%% file: Optimal_trading_using_signals.tex
\documentclass[12pt]{article}

\usepackage[english]{babel}
\usepackage[T1]{fontenc}
\usepackage[applemac]{inputenc}
\usepackage{lmodern}
\usepackage{microtype}
\usepackage{bbm}
\usepackage{tikz}
\usepackage{graphicx}
\usepackage{subfloat}
\usepackage{amsmath}
\usepackage{amssymb}
\usepackage{amsthm}
\usepackage{latexsym}
\usepackage{color}
\usepackage{dsfont}
\usepackage{appendix}

\usepackage{hyperref}
\usepackage{float}
\usepackage{enumitem}
\usepackage{mathabx}
\usepackage{wrapfig}
\usepackage{graphicx} % Required for including images
\graphicspath{{figures/}}

\DeclareSymbolFont{calletters}{OMS}{cmsy}{m}{n}
\DeclareSymbolFontAlphabet{\mathcal}{calletters}
\def\be{\begin{eqnarray}}
\def\ee{\end{eqnarray}}

\def\b*{\begin{eqnarray*}}
\def\e*{\end{eqnarray*}}

\def \E{\mathbb{E}}

\def \P{{\mathbb P}}

\def \R{\mathbb{R}}

\def \eps{\varepsilon}

\def\Fc{{\cal F}}

\def \Fb{\bar{F}}

\def\Qb{{\bar Q}}

\def\pb{{\overline{p}}}

\def\Qt{{\widetilde Q}}

\def \Ctn{{\rm{C}}}

\def\no{\noindent}
\def\x{\times}

\def\={\;=\;}
\def\.{\;.}

\def\eps{\varepsilon}

\def \1{{\bf 1}}
\def \ep{\hbox{ }\hfill{ ${\cal t}$~\hspace{-5.1mm}~${\cal u}$   } }

\def \proof{{\noindent \bf Proof. }}
\def \ep{\hbox{ }\hfill$\Box$}

 \def\normeL2#1{\left\|{#1}\right\|_{L^2}}
\title{Optimal trading using signals}

\author{Hadrien De March\thanks{CMAP, \'Ecole Polytechnique, hadrien.de-march@polytechnique.org.} \and Charles-Albert Lehalle\thanks{Capital Fund Management (Paris) and Imperial College (London).}}

\date{\today}

\begin{document}

\maketitle

\newtheorem{Theorem}{Theorem}[section]
\newtheorem{Lemma}[Theorem]{Lemma}
\newtheorem{Corollary}[Theorem]{Corollary}
\newtheorem{Proposition}[Theorem]{Proposition}
\newtheorem{Remark}[Theorem]{Remark}
\newtheorem{Example}[Theorem]{Example}
\newtheorem{Definition}[Theorem]{Definition}
\newtheorem{Assumption}[Theorem]{Assumption}

%\tableofcontents

\abstract{%
In this paper we propose a mathematical framework to address the uncertainty emerging when the designer of a trading algorithm uses a threshold on a signal as a control.
We rely on a theorem by Benveniste and Priouret to deduce our Inventory Asymptotic Behaviour (IAB) Theorem giving the full distribution of the inventory at any point in time for a well formulated time continuous version of the trading algorithm.

Since this is the first time a paper proposes to address the uncertainty linked to the use of a threshold on a signal for trading, we give some structural elements about the kind of signals that are using in execution. Then we show how to control this uncertainty for a given cost function. There is no closed form solution to this control, hence we propose several approximation schemes and compare their performances. 

Moreover, we explain how to apply the IAB Theorem to any trading algorithm driven by a trading speed. It is not needed to control the uncertainty due to the thresholding of a signal to exploit the IAB Theorem; it can be applied ex-post to any traditional trading algorithm.
\\[1ex]

}

\paragraph{Acknowledgements.} Authors thank Mihail Vladkov, Robert Almgren, Reza Gholizadeh, Isaac Carruthers, and Shankar Narayanan for very fruitful discussions about the modelling of a trading algorithm using a threshold on a signal.

%\tableofcontents

\section{Introduction}
\input{intro.tex}

% market impact and orderflow

\subsection{Modeling optimal trading in automated markets}\label{subsect:model}

In this paper we propose a model at a small time scale (i.e. the one of orderbook dynamics), allowing to design optimal trading strategies using signals.

First of all, one has to note that, at the scale of orderbook dynamics, the time is discrete and event-driven.
Following the literature (like \cite{citeulike:8531765} or \cite{eisler2012price}), we will adopt an event driven model, were an ``event'' can be a trade, or the insertion, the modification or the cancellation of an order at a ``meaningful distance'' of the mid-price (being defined as the middle between the best ask and the best bid price). 
The exposed framework will not address a collection of electronic orderbooks independently (as in \cite{citeulike:10160160}), but will trade in front of one aggregated liquidity pool.

This paper focuses on the optimal behaviour of one large trader. Our methodology can address the cases of a market maker, a proprietary trader or a broker executing the large order of an institutional investor.
To keep the notations and the explanations simple, we will treat the case of a broker trading algorithm.
In such a case the contract is straightforward: buy (or sell) $Q^*$ shares from $t=0$ to $t=T$ at the best possible price. If it is a buy order, the trader is not allowed to sell, and if it is a sell order, he is not allowed to buy.
\medskip

The simplest way to express optimal trading in our framework is as follow:
\begin{itemize}
\item The trading takes place between time 0 and time $T$;
\item During this time interval, the trader will "wake-up" $N$ times (the last one being at $T$);
\item Time steps (i.e. wake-ups) are indexed by $n$, it can be any unit of time (physical time or event time);
\item At each step $n\in \left[| 0,N\right|] $, the price of the stock is $P_{n}$, 
	the inventory of the trader has a size $Q_n$, and the money accumulated by the trader is $M_n$;
\item Each time the trader wakes up, he can read a signal $S_n$ and use it to decide to trade or not immediately at price $P_n$,
	the ``\emph{expected future price change}'' is monotonous in $S$. It means the largest $S$, the more chance to see a price increase in a short future (i.e. before the next weak up);
\end{itemize}

A natural way for the trader to take a decision at step $n$ is to buy only if the signal is larger than a threshold, and to sell when it is lower than another threshold.
The dynamics of the system hence is given by
\begin{equation}\label{eq:dyn:prop}
\left\{ \begin{array}{lcl}
P_{n+1}&=&P_{n} + \Delta P_{n+1}\\
Q_{n+1}&=&Q_{n} + \mathbf{1}_{S_{n}>\theta_n} \\
M_{n+1}&=&M_{n} - \mathbf{1}_{S_{n}>\theta_n}\cdot P_{n}.\\
\end{array}\right.
\end{equation}
Where $X_{n} = (P_{n},Q_{n},M_{n})$ is the state, $\theta_n$ is a \emph{threshold}, chosen just before $t=n$, to be applied to a signal $S$.
We simplify the notation here. We use $S_n>\theta_n$ to express that the threshold $\theta_n$  chosen at $t_n$ is crossed after $t_n$ and before $t_{n+1}$. Remember that $t_n$ are wake-up times of around few seconds to few minutes; as a consequence multiple orderbook events occur between two wake-ups.

The trader targets to find a $\theta_n$ that optimizes a utility function of the final money and remaining quantity : $u(M_{N},Q_{N})$. He basically wants to maximize the money while having $Q_{N}$ as close as possible to a terminal target $Q^{*}$. The control used by the trader to reach this optimum is the sequence of thresholds $\theta_n$.

\subsection{The signal}

Here we will consider a short term predictive signal. 
As an example we will use the signal studied in \cite{lipt13sig}, i.e. the quantity at the best bid divided by the sum of the quantities at the best bid and asks. As underlined in the paper, this signal has more predictive power when the tick is \emph{large}, in the sense the quantities at the best bid  and ask describe better the short term liquidity in the orderbook on large ticks\footnote{An instrument is usually considered as \emph{large tick} when its average bid-ask spread is lower than 1.2 ticks \cite{huang2016predict}.}.
For small tick instrument, it is probably needed to aggregate smartly several limits to obtain the same predictive power.

\begin{figure}[!ht]
   \centering
    \includegraphics[width=.45\textwidth]{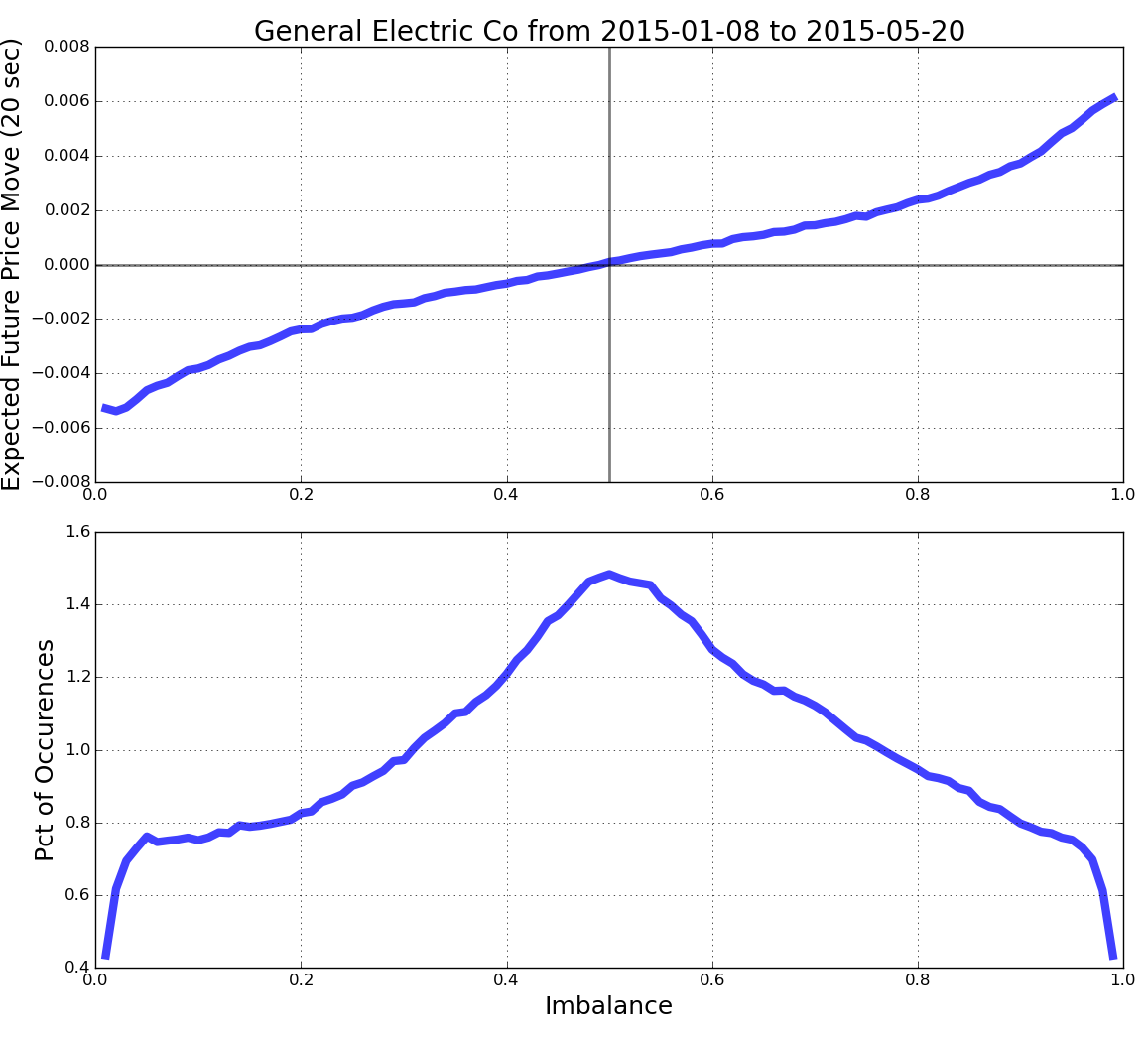}
    \includegraphics[width=.45\textwidth]{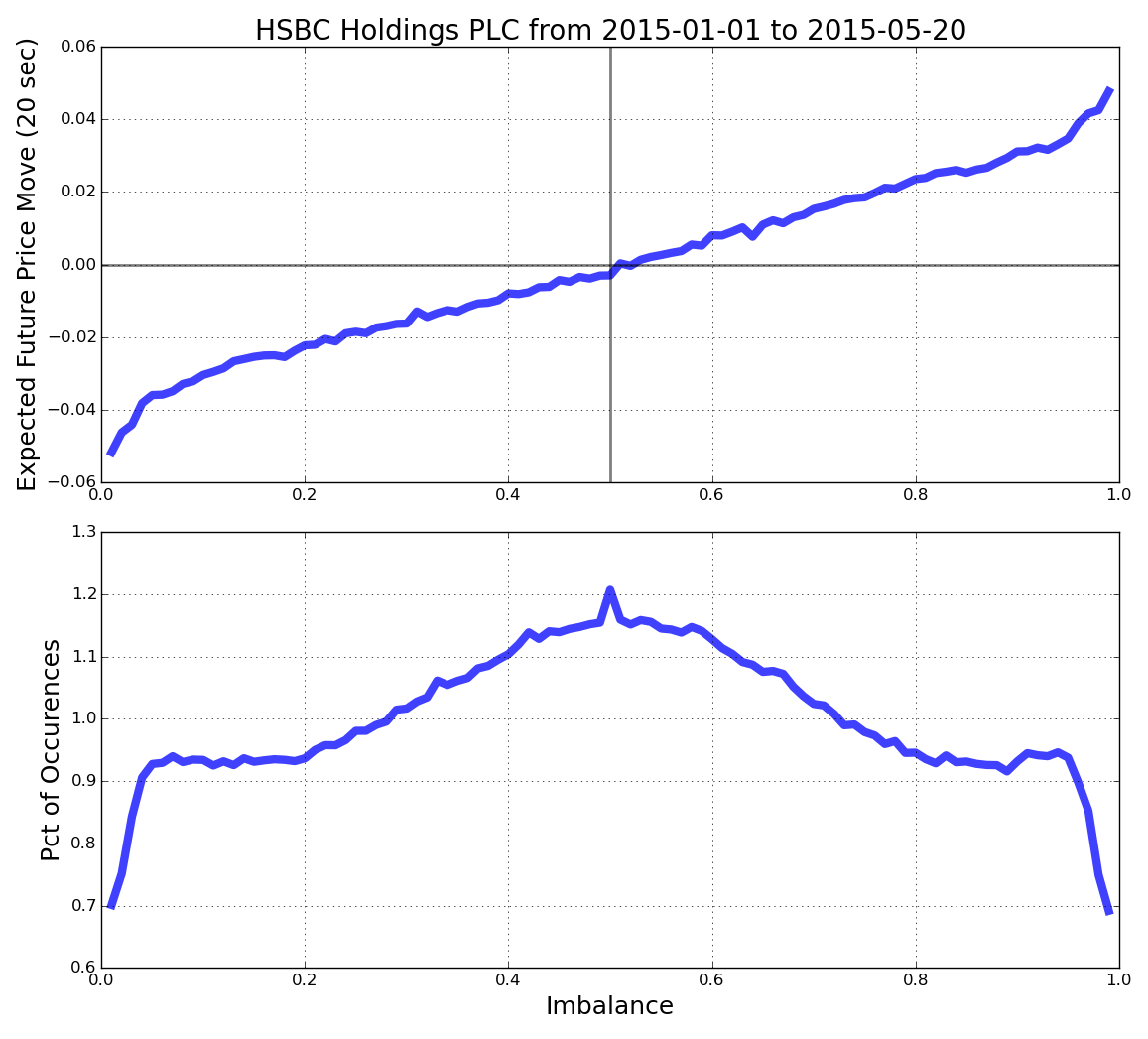}
      \caption{The "\emph{imbalance signal}" for General Electric (left, using 100 million of observations) and HSBC (right, using 590 million of observations). 
      Top: the expected price change after 20 seconds;
      Bottom: the density $f$ of the imbalances values (sampled every update of first bid and ask price or quantity).}
   \label{fig:lipton:sig}
 \end{figure}

We computed the signal of this paper using Capital Fund Management tick-by-tick datasets. Figure \ref{fig:lipton:sig} shows the relation between the expected price move in the future 20 seconds and the imbalance is increasing. Sampled every update of the first bid and ask (quantity or price), the density is rather U-shaped. Major trading venues available are used and aggregated\footnote{We refer to fidessa's \emph{fragulator} to select the venues: Nyse, BATS, NASDAQ and Nyse Arca for these two stocks.}. 

To make it simple we will consider that this signal $S_n$ is i.i.d. and that its interaction with the price has a correlation with the future price change $\Delta P_{n+1}$. As we will see later, an essential notion for solving this optimal trading problem is the gain associated to a quantile of level $p$:
$$
g(p) = \E(\Delta P_{n+1}\cdot \mathbf{1}_{S_n\geq \vartheta(p)})
$$
Note that the quantile $p$ is the probability to obtain a trade when setting the threshold to $\vartheta(p)$. By definition of $p$ and $\vartheta(\cdot)$, we have
$p = \P(S_n\geq \vartheta(p))$.
%As $q(\cdot)$ is a decreasing function, we can express the threshold as a function of the quantile $q$ and then we will also write the gain as a function $g(q)$ of $q$.\\
Finally, we will assume that the signal $S_n$ is a well designed signal, which means that the absolute gain $G(p) = \E\big(\Delta P_{n+1}|S_n\geq \vartheta(p)\big) = g(p) / p$ is an increasing function. 
% As we have the identity $g(p) = G(p)\cdot p$, we conclude that a well designed gain function satisfies $p\mapsto g(p)/p$ is decreasing.
\medskip

\begin{figure}[!ht]
   \centering
	\includegraphics[width=.45\textwidth]{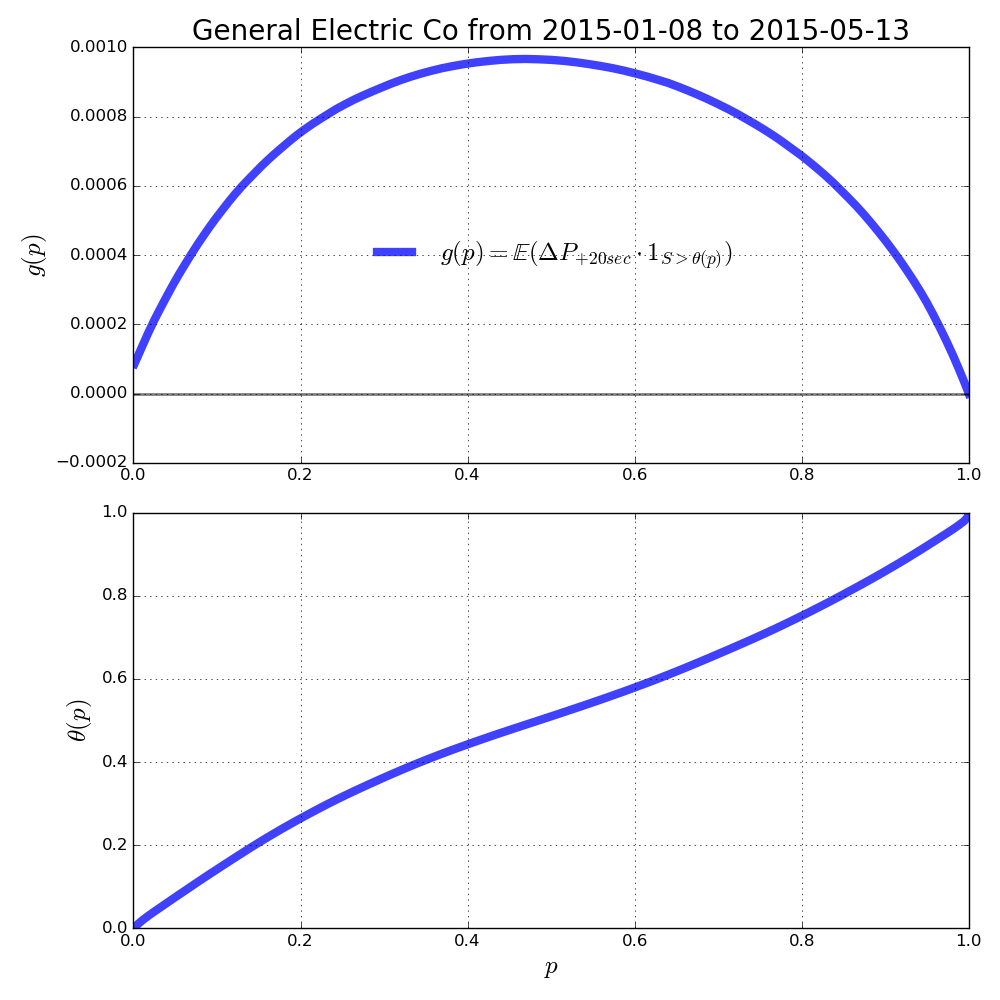}
    \includegraphics[width=.45\textwidth]{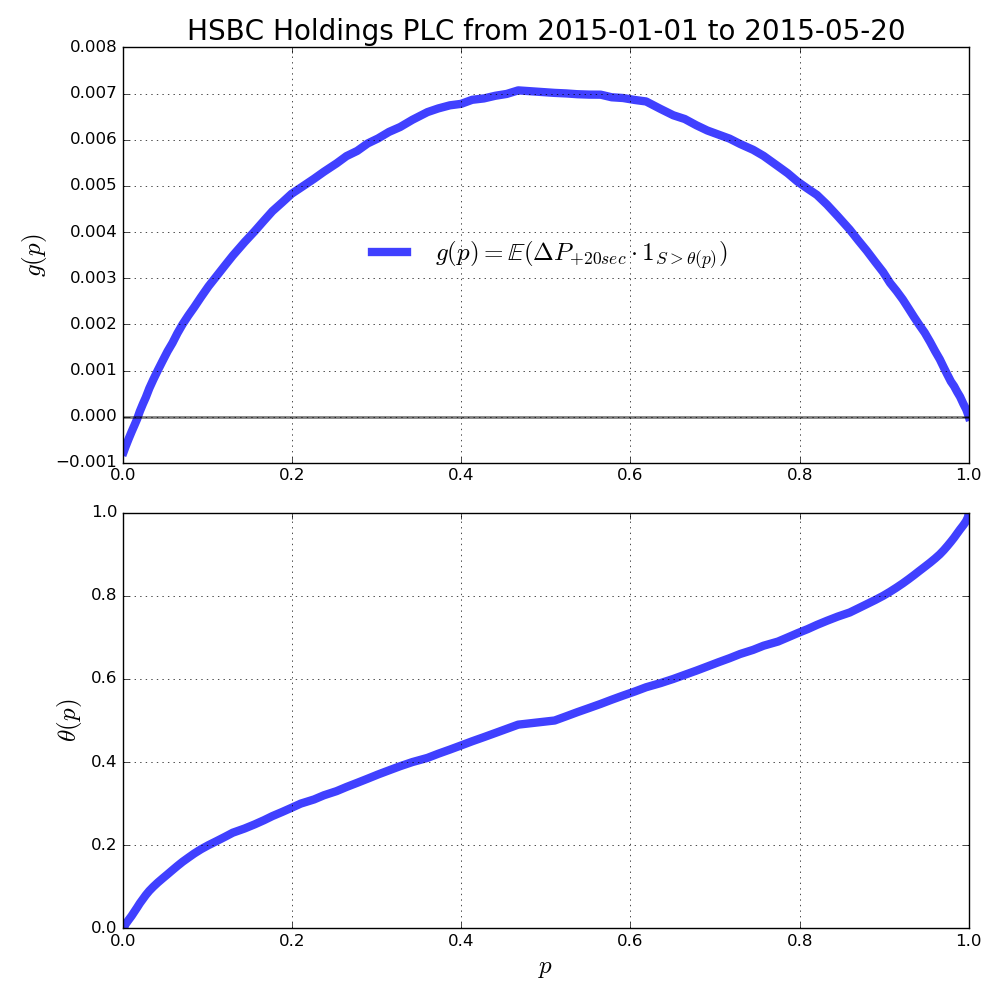}
    \caption{The function $g(p)$ and $\Theta(p)$ for the "\emph{imbalance signal}" on General Electric (left) and HSBC (right).}
   \label{fig:lipton:qgQ}
 \end{figure}

Figure \ref{fig:lipton:qgQ} shows the shape of the different functions linked to the imbalance on real data. 
One can read on the bottom charts: the largest threshold $\vartheta(p)$, the better expected price move, but the lower $p$.
This is a \emph{saturation mechanism}: the more the trader wants to exploit the signal, the less attractive opportunities he has to catch, progressively killing the efficiency of his strategy.

%===================================================================

\section{Trading with a threshold on a signal}

\subsection{Discrete formulation of the problem and the associated uncertainty}

We study the following problem: a broker wants to buy a given quantity $Q^*$ of shares (or contracts). To do that, he will have $N$ opportunities to buy shares during the day (or a given period of time). For each opportunity $n$ from $0$ to $N$, we introduce the following quantities: $Q_n$ is the number of shares already bought, $P_n$ is the stock price, and $M_n$ is the bank account (i.e. the money left). At each time we also observe a short term signal $S_n$ that has a predictive power on the evolution of the price between $n$ and $n+1$. Our control strategy will be to buy one share\footnote{We take a unit of one share for each opportunity, but it can be any number of shares, even random, as far as it is not correlated with the signal nor the price, without changing the result. If the number of shares is correlated with the signal or the price (it can be the case), then other terms will appear in the equations.} each time $S_n$ is higher than a threshold $\theta_n$ that we will determine\footnote{The reader can consider another way to use a signal: maintain one limit order in the book except if the signal is lower than a threshold. The formula will look the same once an adequate change of variable will have been done.}. We recall (\ref{eq:dyn:prop}):
\b*
P_{n+1}-P_n &=&\Delta P_{n+1}\\
Q_{n+1}-Q_n &=& \mathbf{1}_{S_n\ge\theta_n}\\
M_{n+1}-M_n &=& -\mathbf{1}_{S_n\ge\theta_n}P_n.
\e*
We add some assumptions.
\begin{Assumption}\label{ass:signal}
\no{\rm (i)} The signals and price moves vector $(S_n,\Delta P_{n+1})$ is i.i.d.

\no{\rm (ii)} The map $\theta\longmapsto \E[\Delta P_{n+1}|S_n\ge \theta]$ is increasing, and $(P_t)_t$ is martingale.

\no{\rm (iii)} The map $\theta\longmapsto \P[S_n\ge \theta]$ is a continuous bijection $I\longrightarrow [0,1]$ for $I\subset \overline{\R}$. We denote $\vartheta$ its reciprocal map.
\end{Assumption}

Thanks to this assumption, we may define for $p\in[0,1]$ the gain map $g(p):=\E[\mathbf{1}_{S_n\ge\vartheta(p)}\Delta P_{n+1}]$. Indeed, this map is independent of $n$ thanks to Assumption \ref{ass:signal} (i). Furthermore, $g$ is concave by (iii) and $g(0) = g(1) = 0$ by (ii). Indeed, $g(0):=\E[\mathbf{1}_{\emptyset}\Delta P_{n+1}] = 0$, $g(1):=\E[\Delta P_{n+1}] = 0$, and $\frac{g(p)-g(p_0)}{p-p_0}=\frac{\E\left[\mathbf{1}_{\Delta P_{n+1}\in[\vartheta(p_0),\vartheta(p)]}\Delta P_{n+1}\right]}{p-p_0}$. Then $\frac{g(p)-g(p_0)}{p-p_0}\in [\vartheta(p_0),\vartheta(p)]$, and therefore by (iii) $g$ is differentiable at $p$ and $g'(p)=\vartheta(p)$. The concavity of $g$ is a consequence of the fact that $\vartheta$ is decreasing.

The problem we study in this paper is the optimal policy $\theta_n$ to apply in order to maximize the expectation of $M_N$ while executing the $Q^*$ orders:
\be\label{pb:optimal_trading}
\max_{(\theta_n)_{0\le n\le N-1},\, Q_N = Q^*} \E[M_N].
\ee
We may reformulate this problem to show that $Q_n$ and $n$ are the only important state variables. Let introduce for all $n$, $X_n := M_n + (Q^*-Q_n)P_n$, and define the probabilities of buying $p_n := \theta^{-1}(\theta_n)$. We have the fundamental evolution equation
\be\label{eq:evolution_X}
X_{n+1}-X_n = \mathbf{1}_{S_n\ge\theta_n}\Delta P_{n+1}+(Q^*-Q_n)\Delta P_{n+1}.
\ee
The expectation of \eqref{eq:evolution_X} gives that $\E[X_{n+1}-X_n|\Fc_n] = g(p_n)$. Then we have $\E[X_N] = X_0+\E\left[\sum_{n=0}^{N-1}g(p_n)\right]$, and therefore the problem \eqref{pb:optimal_trading} may be rewritten
\be\label{pb:optimal_trading_reformulated}
\max_{(\theta_n)_{0\le n\le N-1},\, Q_N = Q^*} \E[M_N]&=&\max_{(p_n)_{0\le n\le N-1},\, Q_N = Q^*} \E[X_N]\nonumber\\
&=& X_0+\max_{(p_n)_{0\le n\le N-1},\, Q_N = Q^*} \E\left[\sum_{n=0}^{N-1}g(p_n)\right].
\ee
We observe in particular that for this optimization problem, the quantity $Q_n$ and the time are the only important state variables.

\subsection{The continuous limit}

In order to get an HJB control equation, it is natural to pass to the continuous time limit. Notice that $\E[Q_{n+1}-Q_n] = p_n$. By adaptativeness of the optimal control and by linearity, we may look for optimal controls under the shape $p_n = F_N(n,Q_n)$ for some $F_N:\{0,...,N-1\}\x\{0,...,Q^*\}\longrightarrow [0,1]$. In order to pass to the limit, we need another assumption.

\begin{Assumption}[A priori shape of the execution probability]\label{ass:approx}
We assume that $F_N$ stems from a macroscopic smooth map $\bar F\in\Ctn^{0,1}\big([0,T]\x[0,\Qb^*],[0,1]\big)$:
\begin{equation}\label{eq:q:approx}
F_N({n},{Q}) = \bar{F}\left(\frac{nT}{N},\frac{Q}{N}\right) + o({1}/{\sqrt{N}}),
\end{equation}
when $N$ goes to infinity, where the $o$ is in probability, uniformly on all $0\leq n\leq N$, and $\Qb^* := \frac{Q^*}{N}$.
\end{Assumption}
A "noise" term on the quantity will appear in the following results, it will be noted with a bar, see equality (\ref{eq:Q:dynas}).
 
Note that after these definitions:
\begin{itemize}
\item we have a way to model an imperfect knowledge on the signal realizations,
\item we have a way to express the model when $N$ is large.
\end{itemize}

We can now provide the asymptotpic dynamics of the inventory.

\begin{Theorem}[Inventory Asymptotic Behaviour  (IAB)]\label{theo:asympt}
Under Assumption \ref{ass:approx}, the size of the inventory $Q_n$ at any step $n$ given by
\begin{equation}\label{eq:Q:dynas}
Q_{n}=N\Qb_{\frac{nT}{N}}  + o(\sqrt{N})
\end{equation}
Where $(\Qb_t)_{0\le t\le T}$ is solution of the stochastic differential equation
\begin{equation}\label{eq:Q:determ}
\left\{\begin{array}{lcl}
d\Qb_t&=&\Fb(t,\Qb_t) \frac{dt}{T}+\sqrt{\frac{\Fb(1-\Fb)}{N}}(t,\Qb_t)\frac{dW_t}{\sqrt{T}}\\
\Qb_0&=&0.
\end{array}\right.
\end{equation}
\end{Theorem}
\proof
This is a straightforward application of the differential equation theorem (Theorem 1 of Chapter 3 in \cite{benveniste2012adaptive}, see Appendix \ref{sec:app:lim:theo}).
We just need to note $\Delta Q_{n+1}$ is bounded, that its expectation is $\E[\Delta Q_{n+1}|\Fc_n] = \bar{F}\left(\frac{nT}{N},\frac{QT}{N}\right) + o({1}/{\sqrt{N}})$, and that its variance is given by $\mathbb{V}(\Delta Q_{n+1}|\Fc_n) = \bar{F}(1-\bar{F})\left(\frac{nT}{N},\frac{Q}{N}\right) + o(1)$.
\ep

\subsection{The associated Hamilton-Jacobi-Bellman equation}

For the sake of simplicity for the notation, we assume now that $T = 1$, up to scaling the time. Then the case $T\neq 1$ is obtained by replacing $t$ by $t/T$ in the equations. Up to relaxing the regularity constraint on $(\pb_t)$, we may rewrite the optimal trading problem \eqref{pb:optimal_trading_reformulated} when $N\gg 1$ as follows:
\be\label{pb:optimal_trading_continuous}
\max_{(\pb_t)_{0\le t\le 1}\subset[0,1],\,\Qb_1 = \Qb^*}\E\left[\int_0^1g(\pb_{s})ds\right],&\mbox{with }\Qb_t := \int_0^{t} \pb_{s} ds + \frac{1}{\sqrt{N}}\int_0^{t} \sqrt{\pb_{s}(1-\pb_{s})} dW_s.
\ee

Let $V_N(t,Q):= \max_{(\pb_s)_{t\le s\le 1}\subset[0,1],\,\Qb_t = Q,\,\Qb_1 = \Qb^*}\E\left[\int_t^Tg(\pb_s)ds\right]$. We are looking for the value seen from zero: $V_N(0,0)$. The bounds on $\pb_t$ give the limit conditions: $V_N(t,\Qb^*) = V_N(1-t,\Qb^*-t) = 0$, by the fact that $g(0)= g(1) = 0$. For $(t,Q)$ such that $\Qb^*-(1-t)\le Q\le\Qb^*$, the dynamic programming principle provides the HJB partial differential equation satisfied by $V_N$:
\be\label{eq:HJB}
-\partial_t V_N=\max_{p\in[0,1]}\left(g(p)+p\partial_Q V_N + \frac{1}{2N}p(1-p)\partial^2_Q V_N\right).
\ee
We use the fact that $N\gg 1$ to solve the HJB equation.
\begin{Proposition}\label{prop:deterministic}
In the deterministic limit $N = \infty$, the function $V_\infty(t,Q) := (1-t)g\left(\frac{\Qb^*-Q}{1-t}\right)$ is the solution of \eqref{eq:HJB} for $\Qb^*-(1-t)\le Q\le\Qb^*$, and the associated optimal control is given by $\pb^{\rm det}(t,Q) := \frac{\Qb^*-Q}{1-t}$.
\end{Proposition}
\proof
The constraint may be included in a Lagrangian. The resolution becomes
\b*
\min_\lambda\max_{(\pb_s)_{t \le s\le 1}\subset[0,1]}\int_t^1 g(\pb_s)ds-\lambda\left(\int_t^{1}\pb_sds-\Qb^*\right).
\e*
by deriving formally with respect to $\pb_s$, we get that $g'(\pb_s) = \lambda$, which means by strict concavity of $g$ that $\pb_s$ is constant. Therefore the only value that guarantees $\Qb_{1} = \Qb^*$ is $\pb_s = \frac{\Qb^*-Q}{{1}-t}$, whence the result.
\ep

\section{Approximate solutions}

\subsection{Performance ratio}

Our goal in this section is to find closed approximate formulas for the optimal threshold policy. In order to determine if the approximation is satisfying, we shall introduce a performance criterion. Thanks to discrete dynamic programming\footnote{We implement a numerical scheme to obtain the optimal solution and we use it as a benchmark, see Appendix \ref{sec:app:discret_resolution}.}, we may compute the optimal strategy and its expected return ${\rm perf}_{optimal}$. Similar, we may compute the expected return ${\rm perf}_{deterministic}$ of the deterministic strategy from Proposition \ref{prop:deterministic}: $\pb_t = \pb^{\rm det}$. Then the performance criterion of a thresholding policy $\pb_t$ will be the ratio of expected return gained from the deterministic strategy to the optimal strategy:
$${\rm performance\,ratio} = \frac{{\rm perf}-{\rm perf}_{deterministic}}{{\rm perf}_{optimal}-{\rm perf}_{deterministic}},$$
where we denote by ${\rm perf}$ the expected return under $\pb_t$.

\subsection{approximate problem}

We may exploit the fact that $\frac{{1}}{2N}\ll 1$ to get an approximate resolution for \eqref{eq:HJB}. We use the approximation $\partial^2_Q V_N\approx \partial^2_Q V_\infty$ in the equation when $N\longrightarrow\infty$. Notice that $\partial^2_Q V_\infty(t,Q)=({1}-t)^{-1}g''\left(\frac{\Qb^*-Q}{{1}-t}\right) = ({1}-t)^{-1}g''\left(\pb_t\right)$. Then the problem to solve becomes
\be\label{eq:approximate}
\widetilde{V}(t,Q) := \max_{(\pb_s)_{t\le s\le {1}}\subset[0,1],\,\int_t^{1}\pb_s\frac{ds}{{1}}=\Qb^*-Q}\E\left[\int_t^{1-\frac{1}{N}}g(\pb_s)+\frac{g''(\pb_s)}{2}\frac{{1}}{N({1}-t)}\pb_s(1-\pb_s)ds\right].
\ee
Notice that Problem \eqref{eq:approximate} would have been ill-posed if the final bound of the integral had been ${1}$, indeed the factor $\frac{{1}}{N({1}-t)}$ is not integrable when $t\longrightarrow {1}$. However, we know that structurally the value of the problem is zero at time $1-\frac{1}{N}$. Then the integral may be stopped at that time, solving the ill-posedness problem.

\begin{Proposition}
In the particular case where $\E[\Delta P_{n+1}|S_n] = a S_n$ for some $a>0$, and $S_n$ follows a uniform law, we have $g(p)=G p(1-p)$ for some $G>0$.
\end{Proposition}
\proof
This follows from elementary calculations left as an exercise for the reader.
\ep\\
We solve Problem \eqref{eq:approximate} in the case of $g(p):=G p(1-p)$ for some $G>0$. The problem becomes
\be\label{eq:particular}
\widetilde{V}(t,Q) &:=& \max_{(\pb_s)_{t\le s\le {1}}\subset[0,1],\,\int_t^{1}\pb_sds=\Qb^*-Q}G\int_t^{1-\frac{1}{N}}\pb_s(1-\pb_s)\big(1-\gamma(s)\big)ds,
\ee
with $\gamma(s) := \frac{{1}}{N({1}-t)}$.

\subsection{Unconstrained approximate solution}\label{subsect:free}

We may solve the problem under the assumption that the solution is the same without the constraint $(\pb_s)_{t\le s\le {1}}\subset[0,1]$. Then as we assumed that the problem was deterministic, we may solve it like a Lagrangian problem. We are looking for $\min_\lambda\max_{(\pb_s)\subset\R_+}\int_t^{1}\pb_s(1-\pb_s)\big(1-\gamma(s)\big)ds-\lambda\left(\int_t^{1}\pb_sds-(\Qb^*-Q)\right)$. By formally deriving with respect to $\pb_s$, we get that $\pb_t = \frac12 -\frac{\lambda}{2\big(1-\gamma(t)\big)}$. We obtain $\lambda$ by ``injecting it into the constraint'', i.e. integrating on the quantity: $\Qb^*-Q = \frac12({1}-t)-\frac{\lambda}{2}\int_t^{1-\frac{1}{N}}\frac{ds}{1-\gamma(s)}$.
It gives
$\pb_t:= \frac12+\left(\frac{1}{{1}-t}\int_t^{1-\frac{1}{N}}\frac{1-\gamma(t)}{1-\gamma(s)}ds\right)^{-1}\left(\frac{\Qb^*-Q}{{1}-t}-\frac12\right)$. Computing the integral, we finally get:
 \b*
\pb_t&=& \frac{1}{2} + \frac{\pb^{\rm det}(t,Q)-\frac{1}{2}}{\left(1+\frac{\ln\big[N({1}-t)-1\big]}{N({1}-t)}\right)\left(1-\frac{{1}}{
N({1}-t)}\right)}.
\e*
where recall that $\pb^{\rm det}(t,Q):=\frac{\Qt^*-Q}{{1}-t}$.

\subsection{Approximate solution under constraint}\label{subsect:border}

The previous control choice has a limitation as the Lagrangian resolution does not take into account the constraint $0\le \pb_t\le 1$. In order to take this feature into account, we solve the following Lagrangian system: 
$$\min_{\mu_t^0,\mu_t^1\ge 0,\lambda}\max_{(\pb_s)\subset\R_+}\int_t^{1}\pb_s(1-\pb_s)\big(1-\gamma(s)\big)ds-\lambda\left(\int_t^{1}\pb_sds-(\Qb^*-Q)\right)-\int_t^{1}\left(\mu_s^0\pb_t+\mu_s^1(1-\pb_t)\right)ds.$$ 
Deriving this Lagrangian with respect to $\pb_t$ gives the equation $(1-2\pb_t)\big(1-\gamma(t)\big)+\lambda = \mu^0_t-\mu^1_t$. Recall that in Lagrangian analysis, the constraint is not active only if the associated coefficient is zero. Therefore $\mu_t^0$ and $\mu_t^1$ cannot be non-zero at the same time and we may study three cases:
\begin{description}
\item[Case 1:]
$\mu^0_t>0$, then $\pb_t = 0$, and $\mu^0_t = \big(1-\gamma(t)\big)-\lambda>0$.
\item[{Case 2:}] 
$\mu^1_t>0$, then $\pb_t = 1$, and $\mu^1_t = -\big(1-\gamma(t)\big)-\lambda>0$.
\item[{Case 3:}] $\mu^0_t=\mu^1_t=0$, then $0 = (1-2\pb_t)\big(1-\gamma(t)\big)-\lambda$, and $-\big(1-\gamma(t)\big)\le \lambda\le \big(1-\gamma(t)\big)$.
\end{description}

Therefore we see that the effect of this constraint depends only on the comparison between $\lambda$ and $\big(1-\gamma(t)\big)$. There are then still three cases depending on the sign of $\lambda$. The case $\lambda = 0$ immediately gives the constant solution $\pb_t = 1/2$. Now we treat the case $\lambda >0$. In such a case there is a limit time $t_0 := 1-\frac{1}{N(1-\lambda)}$ such that $\pb_t = 0$ if $t\ge t_0$, and $\mu^0_t = \mu^1_t = 0$ if $t\le t_0$. It allows to obtain the following equation on $\lambda$ by using the fact that $\Qb^*-Q = \int_t^{t_0}\pb_sds$:
\b*\label{eq:lambda}
1-2\pb^{\rm det}-\frac{{1}}{N({1}-t)}=\lambda\left(1+\frac{{1}}{N({1}-t)}\ln\left[N({1}-t)\frac{1-\lambda}{\lambda}\right]\right),
\e*
where we denote $\pb^{\rm det}:=\pb^{\rm det}(t,Q) = \frac{\Qb^*-Q}{{1}-t}$. Therefore we get
\b*
\pb_t = 1/2 - \frac{ \lambda/2}{1-\frac{{1}}{
N({1}-t)}}.
\e*

The case $\lambda<0$ stems from the symmetry $\pb^{\rm det}\longrightarrow 1-\pb^{\rm det}$, and $\pb_t\longrightarrow 1-\pb_t$, see Proposition \ref{prop:symmetry}. Then for all $\lambda\in\R$ we have
\b*
\pb_t = 1/2 -\frac{ \lambda/2}{1-\frac{{1}}{
N({1}-t)}},
\e*
with 
\b*\label{eq:lambda_gen}
1-2\pb^{\rm det}-{\rm sign}(1-2\pb^{\rm det})\frac{{1}}{N({1}-t)}=\lambda\left(1+\frac{{1}}{N({1}-t)}\ln\left[N({1}-t)\frac{1-|\lambda|}{|\lambda|}\right]\right).
\e*
This value may be found quickly thanks to a solver, with the initial guess $\lambda_0 := 1-2\,\pb^{\rm det}(t,Q)$. We may also iterate the map $f:x\longmapsto \frac{1-2\pb^{\rm det}-{\rm sign}(1-2\pb^{\rm det})\frac{{1}}{N({1}-t)}}{1+\frac{{1}}{N({1}-t)}\ln\left[N({1}-t)\frac{1-|x|}{|x|}\right]}$ on $\lambda_0$. Iterating 4 times is enough in practice.

\subsection{A systematic method to dramatically improve the performance of the approximation for symmetric gain functions}
\label{subsect:unconstr}
\begin{figure}[H]
\centering
  \includegraphics[width=.99\linewidth]{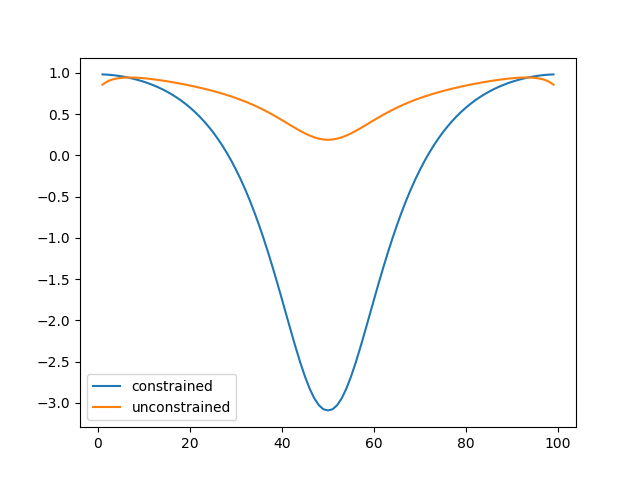}
\caption{For $N=100$; on the $x$-axis: the number of shares (or contracts) to buy (or sell) $Q^*$, and on the $y$-axis: the performance ratios of the approximations of the optimal threshold policy obtained by approximation of Section \ref{subsect:border} (constrained) and Section \ref{subsect:unconstr} (unconstrained).}
\label{fig:performances_base}
\end{figure}

The performance of the previous strategies is not really satisfactory, see Figure \ref{fig:performances_base} (we explain in Section \ref{subsect:perf} how the performance is computed). Indeed the strategy not taking the border into account has an average $60\%$ performance, and the strategy taking the constraint of Section \ref{subsect:border} %borders
into account has a better performance close to the borders, but gets a very negative performance far from the borders. We observe by looking at the numbers that the mismatch in terms of optimal $p$ are strong for small remaining steps and tends to vanish when $N-n$ goes to infinity. There may be several interpretations for this mismatch. The change of performance is very small and then it is very sensitive to the change of $\partial^2_Q V_N$ which is due to the loss of gain, itself due to the term $\partial^2_Q V_N$. The problem is that the formulas that we get involve finding primitives to functions of the shape ${1}/({\ln(x)+x+1})$, which will not generate closed formula.

Furthermore, notice that in all those cases, the optimal $\pb$ does not satisfy Assumption \ref{ass:approx}, as the underlying function $\overline{F}$ does not have a continuous $Q-$derivative at $t = 1$. This explains why most mismatch is located around $t = 1$, therefore we provide a technique that repairs the matching at the final times.

A way to deal with this problem simply consists in shifting the loss function $\gamma$. We replace it by a weakened function $\gamma_\tau(t) := \gamma(t+\tau)$. The interest of simply shifting is that the optimization equations from Section \ref{subsect:free} and Section \ref{subsect:border} bring to the same solution up to a shifting, and therefore generates closed formulas. Now we determine $\tau$ by calibration. We select $\tau$ so that we get the right value for $N-n = 3$. We can get this value from the discrete dynamic programming principle: $p^*_{{1}(1-3/N)}\left(\Qb^*-{1}/N\right) = \frac38$, see Appendix \ref{sec:app:discret_resolution}. It is enough to do it for $\pb^{\rm det} = \frac{1}{3}$ by symmetry with respect to the line $(\pb^{\rm det} = \frac12)$.

For the unconstrained problem we get the equation $\frac38 = \frac12 - \frac{\frac13-\frac12}{\left(1+\frac{2+N\tau}{3}\right)\left(1-\frac{1}{3+N\tau}\right)}$, that leads to the equation $\left(1+\frac{2+N\tau}{3}\right)\left(1-\frac{1}{3+N\tau}\right)=4$. A python solver gives the approximate solution $\tau_{unco} \approx 3.5723/N$. We get the solution:
 \begin{equation}
 \label{eq:unconstrained}
\pb_t= 1/2 + \frac{\pb^{\rm det}(t,Q)-1/2}{\left(1+\frac{\ln\left[N({1}-t)+2.5723\right]}{N({1}-t)}\right)\left(1-\frac{{1}}{
N({1}-t)+3.5723}\right)}.
\end{equation}

For the constrained problem we get the equation $\frac38 = \frac12 + \frac{\lambda}{2\left(1-\frac{1}{3+N\tau}\right)}$, that leads to $\lambda = \frac{\left(\frac{1}{3+N\tau}-1\right)}{4}$ injecting in the equation $\lambda\left(1+\frac{\ln\left((2+N\tau)\frac{1-\lambda}{\lambda}\right)}{3}\right)=1-\frac23-\frac{1+\tau(1-\lambda)}{3}$, we finally get $(2+N\tau)\left(1+\frac{\ln(10+3N\tau)}{3}\right)-N\tau\frac{10+3N\tau}{3}$. A python solver gives the approximate solution $\tau_{const} \approx 1.3445/N$. We get the  solution:
\begin{equation}
\label{eq:constrained}
\pb_t = 1/2 - \frac{ \lambda/2}{1-\frac{{1}}{
N({1}-t)+1.3445 }},
\end{equation}
with 
\b*%\label{eq:lambda_gen}
1-2\pb^{\rm det}-{\rm sign}(1-2\pb^{\rm det})\frac{{1}}{N({1}-t)}\big(1+1.3445 \cdot (1-|\lambda|)\big)\\
=\lambda\left(1+\frac{{1}}{N({1}-t)}\ln\left[\big(N({1}-t)+1.3445\big)\frac{1-|\lambda|}{|\lambda|}\right]\right).
\e*
This value may be computed quickly thanks to a solver, with the initial guess $\lambda_0 := 1-2\pb^{\rm det}(t,Q)$. We may also iterate the map $f:x\longmapsto \frac{1-2\pb^{\rm det}-{\rm sign}(1-2\pb^{\rm det})\frac{{1}}{N({1}-t)}\big(1+1.3445 {1}(1-|x|)\big)}{1+\frac{{1}}{N({1}-t)}\ln\left[N({1}-t)\frac{1-|x|}{|x|}\right]}$ on $\lambda_0$. Iterating 4 times is enough in practice.

\subsection{Comparing performances}\label{subsect:perf}

\begin{figure}[H]
\centering
  \includegraphics[width=.99\linewidth]{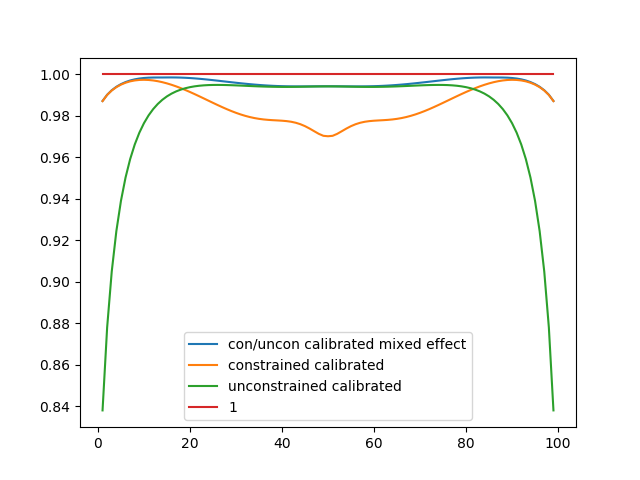}
\caption{For $N=100$; on the $x$-axis: the number of shares (or contracts) to buy (or sell) $Q^*$, and on the $y$-axis: the performance ratios of the approximations of the optimal threshold policy with different approximation schemes: ``\emph{constrained calibrated}'' corresponds to the equation (\ref{eq:unconstrained}); ``\emph{constrained calibrated}'' corresponds to the equation (\ref{eq:constrained}); and ``\emph{con/uncon calibrated mixed effect}'' corresponds to the heuristic exposed in Section \ref{subsect:perf} mixing the two previous solutions.}
\label{fig:performances}
\end{figure}

If we compare for $N = 100$ the performance ratios of these two strategies. Figure \ref{fig:performances} gives this ratio as a function of $Q^*$.

We observe on Figure \ref{fig:performances} that the performance of the shifted strategy taking into account the border effect (``\emph{constrained calibrated}'') is quite good close to the borders. Furthermore, the performance of the shifted strategy not taking the border effect into account (``\emph{unconstrained calibrated}'') is bad close to the borders, but good close to the center. A heuristic strategy consists in taking the border effects into account close to the borders ($1/2-|\pb^{\rm det}-1/2|<0.15$), and stop taking them into account close to the center ($1/2-|\pb^{\rm det}-1/2|\ge 0.15$) this gives the ``\emph{con/uncon calibrated mixed effect}'' curve. The choice of $0.15$ is arbitrary. This mixed-effect strategy gets back at least $98\%$ of the performance available, even $99\%$ out of the extreme $10\%$.

%\clearpage
\section{Application of the Asymptotic Behaviour of the Inventory on other trading strategies}

\input{reuse.tex}

\input{appendix.tex}

\bibliographystyle{plain}
\bibliography{mabib}
\end{document}

%% file: intro.tex
With fragmentation and electronification of markets, the trading process needs to be automated. Trading algorithms can be used to buy or sell a large amount of shares, to route across different trading venue taking profit of the most liquid ones, or to make the markets (see \cite{charles2018market} for more details). 
Academic literature addressed optimal trading from different angles the last twenty years, but few attention has been paid to the use of ``signals'' during the trading process. 
Focus has been made on risk management, since it is one of the usual stakes of mathematical finance.

The optimal trading (or optimal liquidation) literature structured itself in the late nineties around two seminal papers:
\emph{Optimal control of execution costs} \cite{BLA98}, focussed on how to minimize the market impact potentially at the portfolio level, taking profit of the correlations between its components; and 
\emph{Optimal execution of portfolio transactions} \cite{OPTEXECAC00}, focussed on the balance between trading slow (to minimize the market impact) and trading fast (to obtain a price close to the one initiated the decision), in a mean-variance framework close to Markowitz' one \cite{citeulike:571949}.
Both established that as far as you are not concerned by high frequency trading, the \emph{trading speed} is an adequate control.

The literature then evolved towards more sophisticated control of the same components mixed more complex ways. The market impact \cite{bacry15mi} and the price variations during the liquidation \cite{alfonsi2016dynamic} kept the first roles. 
\smallskip

A notable exception is the extension of optimal trading to market making. Following an adaptation of \cite{ho1983dynamics} to modern stochastic control by \cite{avst08}, \cite{Gueant2013Dealing} and \cite{Bayraktar2012Liquidation} simultaneously proposed solutions for different criteria, i.e. adding the intensity of market orders consuming liquidity as a state variable, and using a ``\emph{posting price}'' and not the \emph{trading speed} as control.
In the optimal market making literature, the goal is not to liquidate a position (i.e. a ``metaorder''), but to provide liquidity on both sides of the market (the bid and ask sides), in the hope to earn the bid-ask spread and to not suffer losses because of adverse selection (usually modeled by a price trend) or too large unexpected variations of the midprice.
\smallskip

Coming back to optimal trading: statistical learning methods with different controls like the fragmentation of the metaorder have been proposed to address faster time scale, like in \cite{Pages2011Optimal} or \cite{Laruelle2013Optimal}.
%More recently, different dynamics have been proposed to model the mid price variation, like Hawkes process in [REF:Alfonsi+Blanc].
\medskip

The common point between existing papers in optimal trading before 2015 is they do not make assumptions on the knowledge the optimal trader can have on the future price moves. In practice, it is known short term traders use \emph{signals} to anticipate the value of the price in few minutes or seconds, using them to make the decision to trade now or to wait few seconds or minutes more.
Recent academic papers started to introduce such signals, like \cite{lipt13sig} and \cite{cartea2014buy}.
In the first paper authors proposed a predictive signal of future midprice and provide a detailed analysis and modeling of this signal.
In the second paper, authors put forward a very natural and elegant way to link their signal to permanent market impact. Thanks to this slight modification of the original Almgren and Chriss framework, they can use orderflow as a signal.
%(see [REF:CJ0] for more explanations). 
Nevertheless the orderflow is a very specific signal. They do not provide a generic way to use any predictor to adjust the trading speed.

It may be worthwhile to underline the initial paper of Bertimas and Lo \cite{BLA98} included an AR(1) signal, generating a linear market impact. In the case of a very simple utility function and specific forms of market impact and signal dynamics, this seminal paper can be seen as a special case of the Cartea and Jaimungal framework.
\medskip

The goal of our paper is to provide a way to include signals in optimal trading strategies at a high frequency, using a threshold on this signal as a control: the trader sends an order if and only if the signal is above this chosen threshold. This is very close to some industrial practices. Moreover, a new type of uncertainty appears when you use such a threshold because it does not guarantee to trade.
Setting a threshold at a level $\vartheta(p)$ corresponding to the quantile of level $p$ of the signal does only guarantee that the probability to obtain a transaction is $1-p$.
We believe it is crucial since it is the most natural way to render the idea of the \emph{saturation of a signal}, used in practice by all traders: it is not possible to exploit any signal as strongly as one wants. 
If the trader targets only the highest quantile of the signal (i.e. the best opportunities), he will have only few occasions to trade, and hence he will probably not finish his order on time.
As a consequence, traders having largest quantities to trade, naturally take less profit of their signal per trade.
In short: the more a trader can afford to be opportunistic, the more free he is to choose high quantile, and hence the best price he will obtain at the end.

Another specificity of this paper is we model the price dynamics in a discrete time framework (the time can be seen as event time or physical time without great change) to be as realistic as possible. One of the key points of our work is to provide a natural approximation of the dynamics in continuous time, allowing to solve it an elegant way while controlling the associated approximation error.

Beside, going back to risk control, we show how our approach can provide to any existing trading strategy using thresholds on a signal a way to approximate the uncertainty due to he randomness of the realized trading speed. Even once a strategy is fixed using any arbitrary criterion, our approximation can provide uncertainty bounds on the traded quantities at any point in time.

Our work largely relies on stochastic approximation results of \cite{benveniste2012adaptive}.

The first section of this paper describe the kind of trading signal we have in mind, providing numerical examples. Section two starts with the discrete formulation of a trading algorithm using a signal, expresses our Inventory Asymptotic Behaviour Theorem, and shows its implication on the asymptotic dynamics of a trading algorithm. Section three is dedicated to solving these asymptotic problem for a specific utility function and Section four applies the IAB Theorem to the soltion of the well-know Almgren-Chriss optimal trading problem.

%% file: reuse.tex
In this section we show how to use our framework to understand, measure and anticipate the uncertainty due to the use of threshold-driven policies in a well-known optimal liquidation framework.

Designers of trading algorithms are confronted to the following dilemma: on the one hand there is now a plethoric literature explaining how to design an algorithm minimizing risk-driven cost functions (see for instance \cite{OPTEXECAC00}, \cite{CJ15trade} or \cite{dang12opt});
on the other hand they would like to use fine tuned signals to drive their trading decisions. To the knowledge of authors, only two papers (\cite{eayl17signal} and \cite{CJ11flows}) propose frameworks to inject a signal in the now "standard" risk-driven frameworks. But their control is the trading speed, and not a threshold to be applied to the signal. The most natural way to use a signal once a desired trading speed is computed is to choose a threshold that is crossed by the signal on average a number of times per unit of time equals to the ``optimal trading speed''. 
Typically the repartition function of the signal can be inverted and this inverse gives a natural relationship between a desired trading speed and a threshold to apply on the signal. This is a practice commonly used by designers of trading algorithms. This paper provides a way to address the uncertainty arising from the fact that, if the signal is meant to cross the threshold ``on average'' the desired number of times, the realizations of the signal not really cross the threshold this number of times.
To really control this uncertainty, the reader has to refer to the previous sections. 

Besides, the framework of this paper allows to anticipate the level of this uncertainty not only for trading algorithms designed with our framework, but also for any trading algorithm. In this section, it is hence no more about being "optimal" with respect to this uncertainty, but simply to quantify this uncertainty when it has not been taken into account.
To illustrate how to do it, this section apply our results, and especially our Inventory Asymptotic Behaviour (IAB) Theorem, to the celebrated Almgren-Chriss framework.

\subsection{Notations for the Almgren-Chriss framework}

We simply rely on \cite{OPTEXECAC00} to formulate the optimal liquidation of a stock within a mean-variance framework.
We mix the notations of the original paper with ours to help the reader to make the connection between the two frameworks:
$T$ is the time to liquidate a position of size $Q^*$, in an arithmetic Brownian market with a volatility $\sigma$ and market impact parameters of $\eta$ for the temporary part and $\gamma$ for the permanent part. The risk aversion of the minimization program is $\lambda$. There is no signal in this framework; the result of a mean-variance optimization over $N$ time steps\footnote{Notice that this constant $N$ from \cite{OPTEXECAC00} is different than the one introduced in Subsection \ref{subsect:model}.} is a deterministic optimal trading speed $\nu_{\tau k}$, constant on the $k$th time step $[k\tau,(k+1)\tau]$, where $\tau=T/N$ is the discretization step:
$$\nu_{\tau k} = 2\frac{\sinh(\kappa \tau/2)}{\sinh(\kappa\tau)}\cdot \frac{Q^*}{\tau}\cdot \cosh\{\kappa\cdot(T-\tau k)\},$$
where $\kappa$ stands for $\tau^{-1}\cosh^{-1}\{1+({\widetilde \kappa}\tau)^2/2\}$ and tilde variables are renormalization of the original ones:
$$\widetilde\kappa=\frac{\lambda\sigma^2}{\widetilde\eta},\quad %
\widetilde\eta=\eta\left(1-\frac{\gamma\tau}{ 2\eta}\right).$$

\subsection{Deducing a threshold policy from a trading speed}

The optimal speed resulting from an optimization (here we develop the example of $\nu_{\tau k}$, but this approach can be used for any other trading speed, obtained for instance in \cite{CJ15trade} or \cite{dang12opt}) is only one component of a trading algorithm.
The next step is to interact with liquidity at the time scale of events in an orderbook. Typically the goal of these interactions is to buy $\nu_{\tau k}\cdot\tau$ shares between time $\tau k$ and time $\tau (k+1)$ at the best price.

In practice, designers of trading algorithms use different means to obtain these $\nu_{\tau k}\cdot\tau$ shares.
They can use a stochastic algorithm, like in \cite{Pages2011Optimal}, to learn online the optimal distance to the mid-point for limit orders, or they can control directly the distance to the mid-point under liquidity consumption models, like in \cite{Gueant2013Dealing}.
We explain here how to use a signal to obtain the desired number of shares. Here a signal is a custom indicator such that it is on average better to buy (respectively to sell) when the signal is high (resp. low). The common practice is to set a threshold $\theta_{\tau k}$ when $t=\tau k$ and to keep it during a time $\tau$, waiting for the signal to cross it. This threshold can also be updated in real time, or according to a stopping time that is $\tau$ or the time at which the signal is crossed if any. Another common practice is that as soon as the signal crosses $\theta_{\tau k}$, the algorithm sends a market order or a limit order, or a cancellation instruction\footnote{A very simple strategy would be, for a buy metaorder (assuming that the higher the signal, the higher the expectation of short term price returns), to send a market order if the signal is very high, to maintain a limit order in the book if the signal is not too negative, and to cancel all orders and wait if the signal is too negative.}. For simplicity we will deal with market orders only in our example.

There is a very simple way to deduce the threshold to apply to a signal once the desired trading speed is computed:
\begin{enumerate}
    \item denote $u$ the average number of observations of the signal during a unit if time\footnote{The total number of observations $N$ of the signal, introduced in Subsection \ref{subsect:model}, is given by $uT$.},
    meaning that the algorithm looks at the signal $u\tau$ times during a typical time interval $\tau$.
    \item Set the threshold $\theta_{\tau k}$ such that it is the value of the quantile with a probability $p_{\tau k}=\frac{\nu_{\tau k}\cdot\tau}{u\tau}$.
    \item Apply the IAB theorem with $F(t,Q)=p_{\tau \left\lfloor\frac{t}{\tau}\right\rfloor}$.
\end{enumerate}
That for it is enough that the probability $p_{\tau k}$ is a function of the (remaining) time and the (remaining) quantity to trade.

For our Almgren-Chriss example, it implies:
$$F(t)=\underbrace{2\frac{\sinh(\kappa \tau/2)}{\sinh(\kappa\tau)}}_{a(\kappa,\tau)}\cdot \frac{Q^*}{ u\tau}\cdot \cosh\left\{\kappa\cdot\left(T-\tau \left\lfloor\frac{t}{\tau}\right\rfloor\right)\right\}.$$
Since our optimal trading speed is deterministic, $F$ is a function of the remaining time only,\footnote{For other frameworks like the Cartea-Jaimungal or the Dang-Bouchard-Lehalle ones, $F$ will be a function of time and of the remaining quantity to trade.} and not of $Q$, the remaining quantity to trade.

\begin{figure}[H]
    \centering
    \includegraphics[width=.8\linewidth]{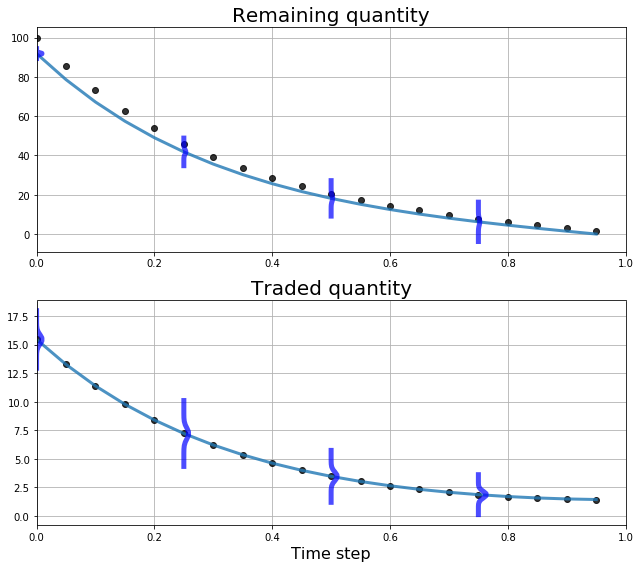}
    \caption{The Discrete Almgren-Chriss result (dots), the continuous version (blue line) and slices of the IAB derived asymptotic distribution for an example ($T=1$, $Q^*=100$, $\eta=1$, $\gamma=1$,  $\sigma=1$, $\lambda=3$). Top panel: Remaining quantity; Bottom Panel: Trading speed.}
    \label{fig:ABI:AC}
\end{figure}

\subsection{A straightforward application of the IAB Theorem}

By the facts that $F$ is continuous and independent of $Q$, Assumption \ref{ass:approx} is satisfied on each interval $[k\tau,(k+1)\tau]$. Injecting $F(t)$ in the IAB Theorem gives immediately the asymptotic distributions when $u\tau \gg 1$ of the realized trading speed ${\widetilde\nu}_{\tau k}:=\frac{{\widetilde Q}_{\tau(k+1)}-{\widetilde Q}_{\tau k}}{\tau}$ and of the number of shares ${\widetilde Q}_{\tau k}$ that will have been really bought at time $\tau k$:
\begin{equation}
\label{sec:uncertain:AC}
{\widetilde\nu}_{\tau k} \sim %
{\cal N}\left(\nu_{\tau k}, %
v(\tau k)\cdot\frac{ u }{\tau}\right), \quad %
{\widetilde Q}_{\tau k}\sim{\cal N}\left( \widetilde{Q}_{\tau k}^{\rm det},  V(\tau k)\cdot uT\right),
\end{equation}
with
$$\widetilde{Q}_{t}^{\rm det}:=Q^*\cdot\frac{\sinh\{\kappa(T-t)\}}{\sinh(\kappa T)},$$
$$v(t):=a(\kappa,\tau) \frac{Q^*}{ u\tau} \cosh\left\{\kappa\cdot\left(T-\tau \left\lfloor\frac{t}{\tau}\right\rfloor\right)\right\} \left(1-a(\kappa,\tau) \frac{Q^*}{ u\tau} \cosh\left\{\kappa\cdot\left(T-\tau \left\lfloor\frac{t}{\tau}\right\rfloor\right)\right\}\right),$$
and  $V(t)=\frac{1}{T}\int_{s=0}^t v(s)\, ds$.

Figure \ref{fig:ABI:AC} shows these distributions for few values of $k$ and their expected values as lines. Their expectations are, by design, the targeted ones, i.e. the continuous version of the Almgren-Chriss framework. Thanks to the IAB Theorem, it is now possible to know an asymptotic of the distribution of these quantities at any point in time, as disclosed by formula (\ref{sec:uncertain:AC}).

%% file: appendix.tex
\appendix
\section*{Appendix}
\section{Limit theorem}
\label{sec:app:lim:theo}

We state the theorem from \cite{benveniste2012adaptive} that we use for the continuous limit.

Let $(\theta_n)_{n\ge 0}$ satisfying for $n\ge 1$:
\b* 
\theta_n = \theta_{n-1}+\gamma H(\theta_{n-1},X_n)+\gamma^2\eps_n(\theta_{n-1},X_n),
\e* 
Where the state has a dynamic Markov representation controlled by $\theta$, hence
\b* 
\P\left[\xi_n\in G|\xi_{n-1},\xi_{n-2},...;\theta_{n-1},\theta_{n-2},...\right]=\int_G\pi_{\theta_{n-1}}(\xi_{n-1},dx)\\
X_n = f(\xi_n),
\e* 
where $\pi_\theta$ is the transition probability of the $\theta-$dependent Markov chain $\xi_n$, and $f$ is a function. We assume the existence of a locally Lipschitz $h(\theta):=\E_\theta\big[H(\theta,X_n)\big]$, so the ODE
\be\label{eq:ODE}
\theta' = h(\theta),&\theta(0)=\theta_0,
\ee
has a unique maximal solution.

\begin{Assumption}[\cite{benveniste2012adaptive}, Chapter 2, Assumption A.1]\label{ass:regul}
There exists a fixed cylinder of diameter $\eta>0$ containing the trajectory $\big(\theta(t)\big)_{0\le t \le T}$ of ODE \eqref{eq:ODE} in which $h$ is locally Lipschitz, and for all compact set $K$, the function $\eps_n(\theta,X)$ is uniformly bounded for $(\theta,X)$ in $K$.
\end{Assumption}

Then the differential equation theorem holds:
\begin{Theorem}[\cite{benveniste2012adaptive}, Chapter 2, Theorem 1]\label{thm:diff_equation}
Let $\eps>0$, and $\gamma>0$ small enough, then under Assumption \ref{ass:regul} we have
\b* 
\P\left[\max_{n:t_n\le T}\|\theta_n-\theta(t_n)\|>\eps\right]\le C(\gamma,T),
\e* 
where, for fixed $T<\infty$, $C(\gamma,T)$ tends to zero as $\gamma$ tends to $0$.
\end{Theorem}

Let $\tilde\theta^\gamma_{t_n} := \gamma^{-\frac12}\big(\theta_n-\theta(t_n)\big)$, and let $\tilde\theta^\gamma_{t}$ denote the trajectory obtained by linear interpolation between the times $t_n$. Then we have the following theorem:

\begin{Theorem}[\cite{benveniste2012adaptive}, Chapter 3, Theorem 1]\label{thm:central_limit}
When $\gamma \longrightarrow 0$, the difference $(\tilde\theta_t^\gamma)_{0\le t\le T}$ converges weakly towards the solution $(\tilde\theta_t^\gamma)_{0\le t\le T}$ of the stochastic differential equation
$$d\tilde\theta_t = \frac{dh}{d\theta}[\theta(t)]\cdot \tilde\theta_t dt + R^{\frac12}[\theta(t)]\cdot W_t,$$
where $(W_t)$ is a standard vector Wiener process, and $R(\theta):=\sum_{n=-\infty}^\infty{\rm cov}_\theta\left[H(\theta,X_n)H(\theta,X_0)\right]$.
\end{Theorem}

\section{Discrete resolution}
\label{sec:app:discret_resolution}

We provide the discrete Hamilton-Jacobi-Bellman equation that solves \eqref{pb:optimal_trading_reformulated}. Let
\b* 
V(n,Q) := \max_{(p_k)_{n\le k\le N-1},\, Q_N = Q^*} \E\left[\sum_{k=n}^{N-1}g(p_k)\right]
\e* 

The result we are looking for is $V(0,0)$. We find it by a backwards resolution. For all $0\le n\le N$, we have $V(n,Q^*) = V(N-n,Q^*-n) = 0$, and the discrete dynamic programming principle gives:
\be\label{eq:prog_dyn}
V(n,Q) &=& \sup_{p\in[0,1]}g(p)+p V(n+1,Q+1)+(1-p)V(n+1,Q).
\ee

If we assume that $g(p) = Gp(1-p)$, we get that $p=\max\left(0,\min\left(1,\frac{\frac{V(n+1,Q+1)-V(n+1,Q)}{G}+1}{2}\right)\right)$. If we have that $|V(n+1,Q+1)-V(n+1,Q)|\le G$, then we have the equation
\b* 
V(n,Q) &=& V(n+1,Q)+\frac14\left(\sqrt{G}+\frac{V(n+1,Q+1)-V(n+1,Q)}{\sqrt{G}}\right)^2. 
\e* 

We may use these formulas to compute $V(N-3,\cdot)$. We have $V(N,Q^*)= V(N-1,Q^*)= V(N-1,Q^*-1) = 0$. Similar $V(N-2,Q^*) = V(N-2,Q^*-2)=0$, and $V(N-2,Q^*-1) = \sup g(p) = \frac14 G$, with $p_{N-2}(Q^*-1)= \frac12$. Finally $V(N-3,Q^*-1) = V(N-3,Q^*-2) = \frac{25}{64}G$, with $p_{N-3}(Q^*-1)= \frac38$ and $p_{N-3}(Q^*-2)= \frac58$.

Finally, we explicit the symmetry relationship verified by $V$.

\begin{Proposition}\label{prop:symmetry}
We assume that for all $p\in[0,1]$, we have $g(1-p) = g(p)$, then for all $n,Q$, we have $V(n,Q) = V\Big(n,Q^*-\big(N-n-(Q^*-Q)\big)\Big)$, and $p_n(Q) = 1-p_n\Big(Q^*-\big(N-n-(Q^*-Q)\big)\Big)$.
\end{Proposition}

Proposition \ref{prop:symmetry} may be proved by backwards induction using \eqref{eq:prog_dyn}.